\documentclass[12pt]{iopart}

\expandafter\let\csname equation*\endcsname\relax
\expandafter\let\csname endequation*\endcsname\relax
\usepackage{amsmath}

\usepackage{hyperref}
\hypersetup{
    colorlinks=true,
    linkcolor=blue,
    filecolor=magenta,      
    urlcolor=cyan,
}

\usepackage{graphicx}
\usepackage{color}
\usepackage{tikz}
\usepackage{soul}
\usepackage{subcaption}
\usepackage{tabularx}
\usepackage{xspace}
\usepackage{cite}

\usepackage{lineno}

\usepackage{pifont}
\usepackage{amssymb}
\usepackage{xfrac}
\newcommand{\cmark}{\checkmark\xspace}
\newcommand{\xmark}{\ding{55}\xspace}%
\newcommand{\xcmark}{\checkmark\,/\,\ding{55}\xspace}%

\newcommand{\red}[1]{#1}

\newcommand{\spectroplot}{
\begin{figure*}
    \centering
    \includegraphics[width=\textwidth]{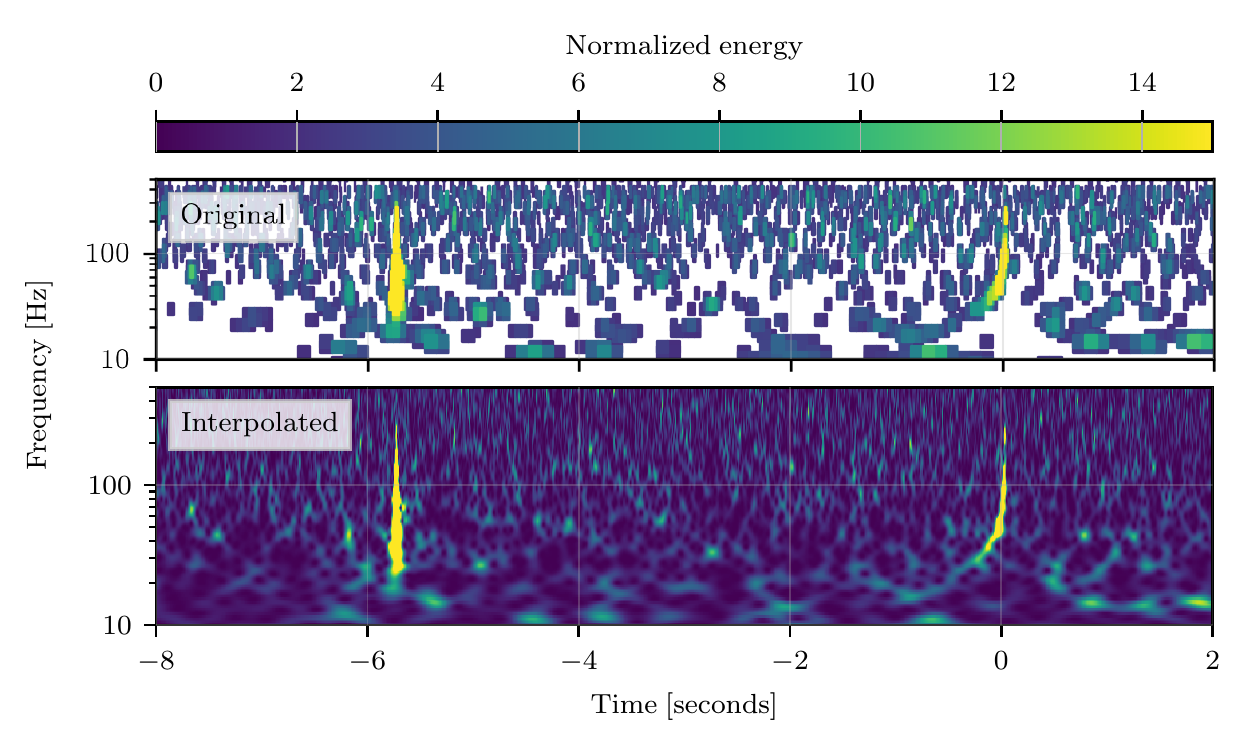}
    \caption{A comparison of a Q-gram (top) and spectrogram (bottom) of the same data around GW170814\cite{GW170814} which shows both a gravitational-wave signal and a glitch. The excess power at $t = 0$ is the actual event, while the glitch is visible at $t = -6$. In both panels, the data was visualized in the time-frequency domain using the Q-transform~\cite{Chatterji_thesis,Chatterji:2004qg,Macleod:2021}. However, the Q-gram shows all tiles with an energy above 2.5 while the spectrogram uses interpolation to estimate the measured energy at all time-frequency points. This statistic used in this work is based on the Q-gram, although interpolated spectrograms similar to this example are also commonly used for data interpretation. }
    \label{fig:spectrogram}
\end{figure*}
}

\newcommand{\distfig}{
\begin{figure*}[t]
    \centering
    \includegraphics[width=\textwidth]{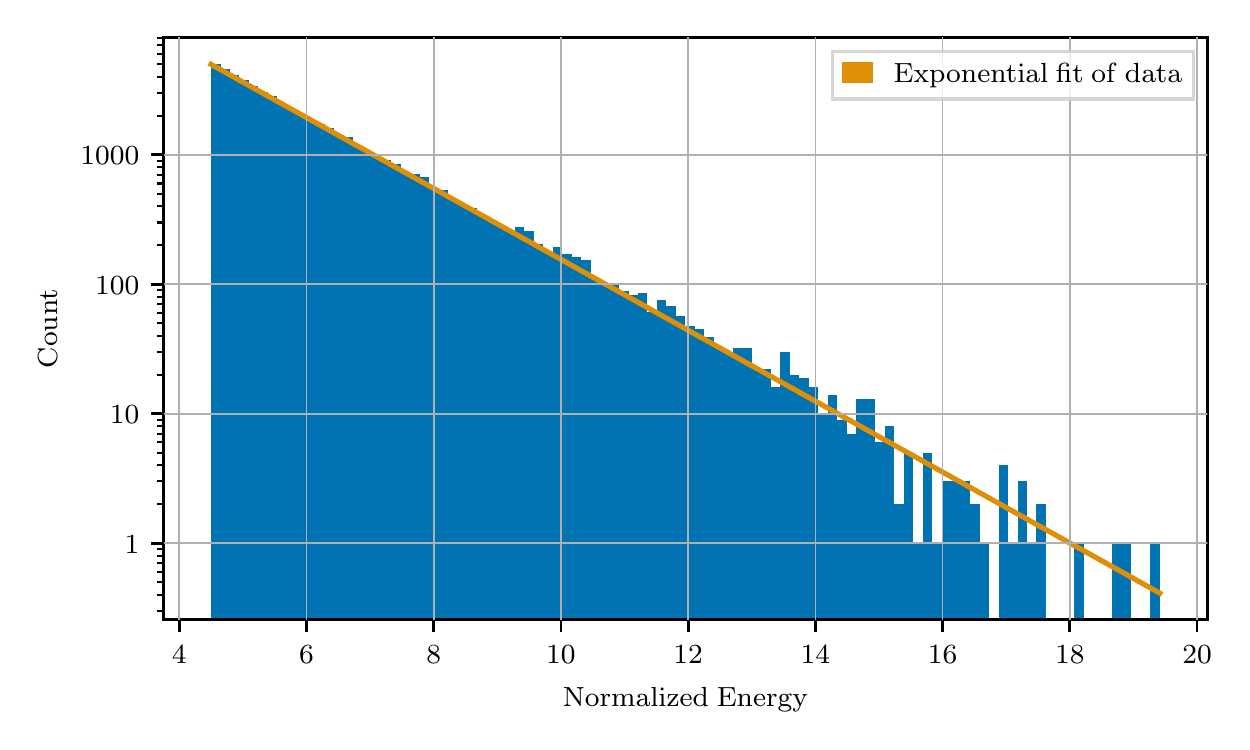}
    \caption{An example of the measured distribution of tile energies in simulated Gaussian noise. The histogram data (blue) is then fit to an exponential (orange). We use this fit as the basis for our p-value calculation. Any tiles with higher energies than would be expected based on this fit are considered significant.}
    \label{fig:distfig}
\end{figure*}
}

\newcommand{\gaussppplot}{
\begin{figure*}[t]
    \centering
    \begin{subfigure}[b]{0.45\textwidth}
    \includegraphics[width = 80mm]{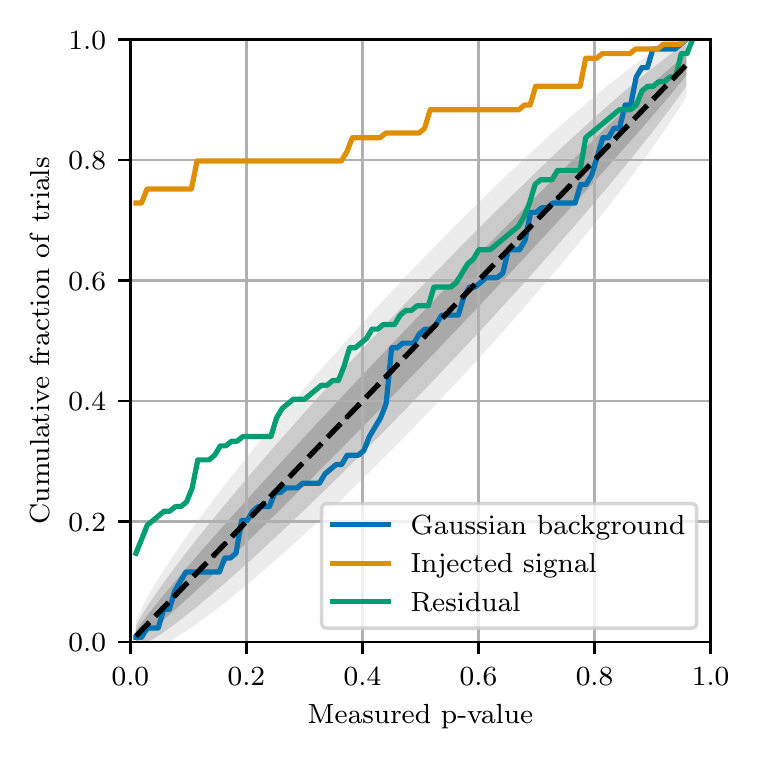}
    \caption{Q fixed}
    \end{subfigure}
    \begin{subfigure}[b]{0.45\textwidth}
        \includegraphics[width = 80mm]{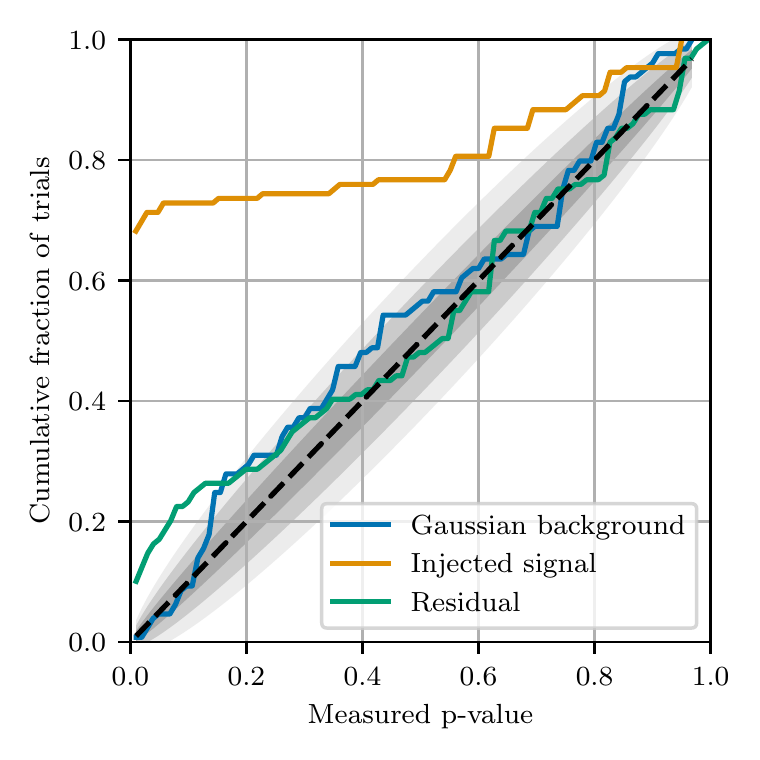}
        \caption{Q varied}
    \end{subfigure}
    \caption{The distribution of the measured p-values from our tests with simulated Gaussian data. The left panel shows the p-values when we fix Q for the tiling in our discretized Q-transform, and the right panel shows the p-value distribution when we test a range of Q values and choose the Q-gram with the highest individual tile energy. The gray regions show the 1, 2, and $3\sigma$ uncertainty in the expected distribution. The distribution of p-values is more consistent with the expected distribution when the Q is fixed.}
    \label{fig:gaussppplot}
\end{figure*}
}

\newcommand{\glitchfig}{
\begin{figure*}[t]
    \centering
    \begin{subfigure}[b]{0.45\textwidth}
    \includegraphics[width = 80mm]{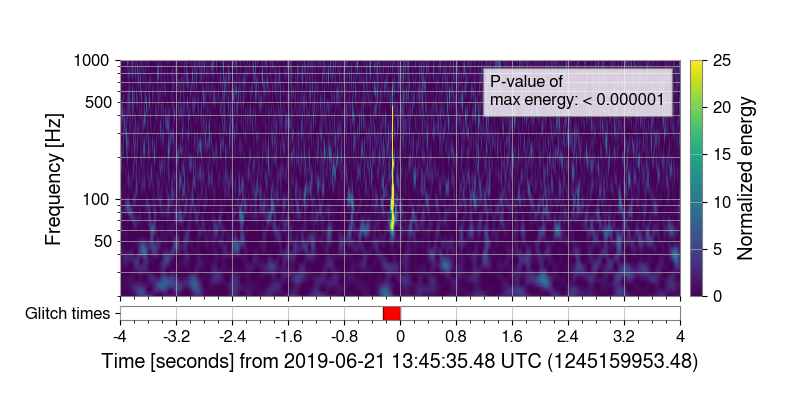}
    \caption{Initial data, glitch only}
    \end{subfigure}
    \hfill
    \begin{subfigure}[b]{0.45\textwidth}
    \centering
    \includegraphics[width = 80mm]{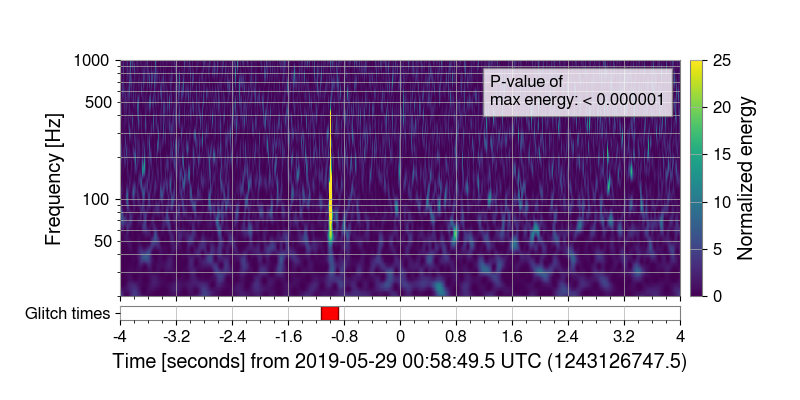}
    \caption{Initial data, glitch only}
    \end{subfigure}
    
    \begin{subfigure}[b]{0.45\textwidth}
    \centering
    \includegraphics[width = 80mm]{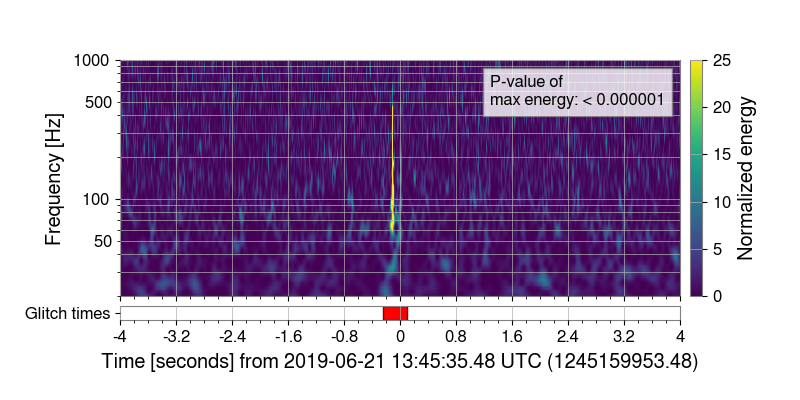}
    \caption{Glitch with injected signal}
    \end{subfigure}
    \hfill
    \begin{subfigure}[b]{0.45\textwidth}
    \centering
    \includegraphics[width = 80mm]{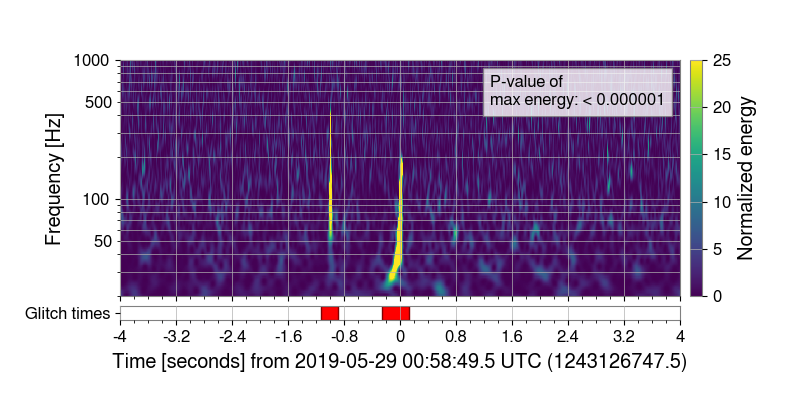}
    \caption{Glitch with injected signal}
    \end{subfigure}
    
    \begin{subfigure}[b]{0.45\textwidth}
    \centering
    \includegraphics[width = 80mm]{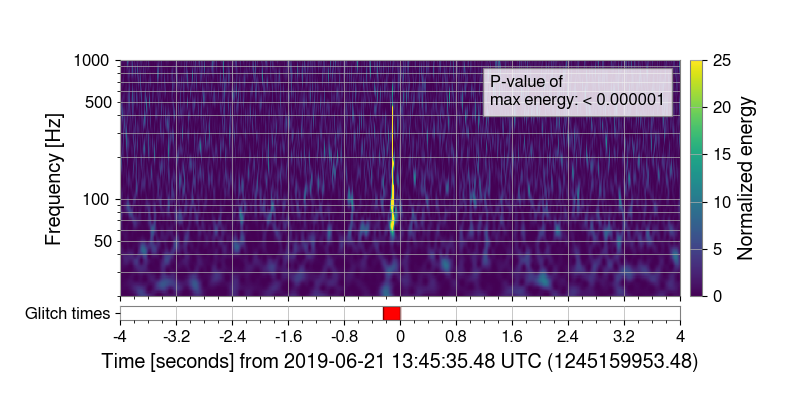}
    \caption{Subtraction Performed}
    \end{subfigure}
    \hfill
    \begin{subfigure}[b]{0.45\textwidth}
    \centering
    \includegraphics[width = 80mm]{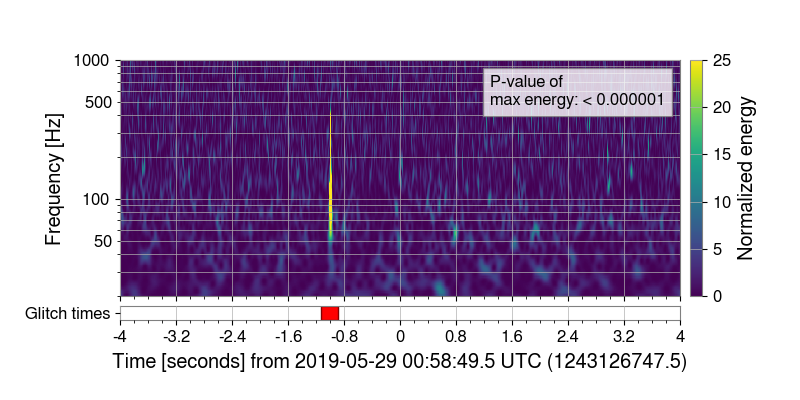}
    \caption{Subtraction Performed}
    \end{subfigure}
    
    \caption{The results of tests performed on data with a loud glitch. The left column shows the matched filtering performed on a smaller window within 0.1 seconds of the injection time, allowing us to successfully subtract an injected signal, even when it occurs contemporaneously with a glitch. Note that we can use the data quality flagging to evaluate if the signal was successfully subtracting even if it is not clear visually. The right column shows the recovery performed with the subtraction allowed during any peak SNR from the whole data section, which works for sufficiently loud gravitational-wave signals, as demonstrated here.}
    \label{fig:glitch}
\end{figure*}
}

\newcommand{\realppplot}{
\begin{figure*}[t]
    \centering
    \includegraphics[width=0.9\textwidth]{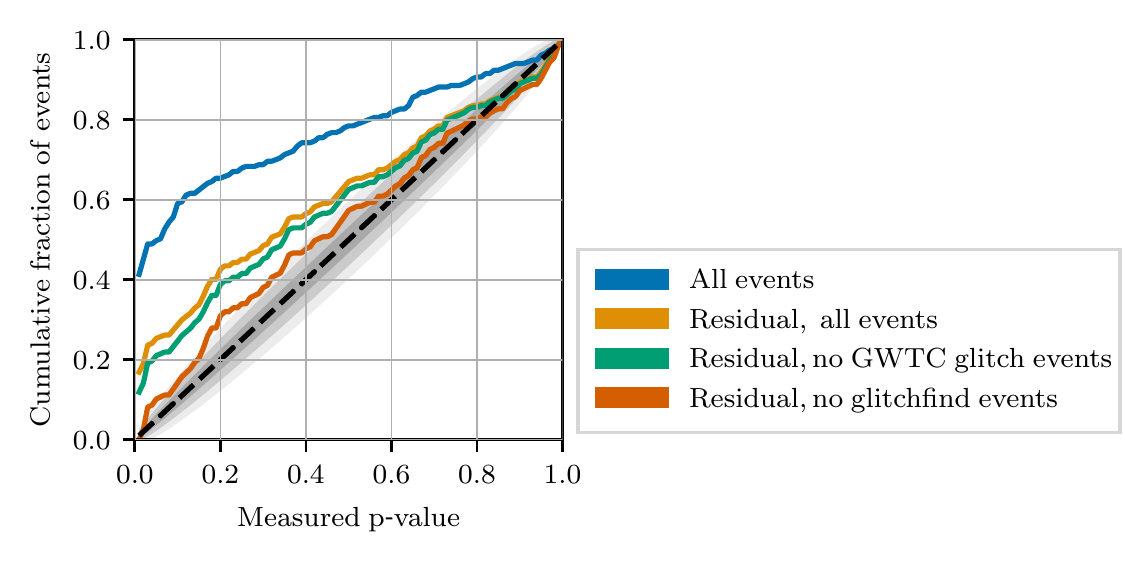}
    \caption{The measured distribution of p-values for our tests with astrophysical events from O1, O2, and O3. We compare the distribution before (blue) and after subtraction of the best fit signal parameters (yellow). We further remove events that are known to contain glitches based either on the reported lists of events that required glitch mitigation (green) or based a p-value threshold of 0.01 (red). 
    The gray regions show the 1, 2, and $3\sigma$ uncertainty in the expected distribution.
    After removing events that required glitch mitigation, we still see an excess of cases with low p-values.
    In comparison, removing events below our p-value threshold results in the p-value distribution being much more consistent with that expected with Gaussian noise. In our figure, ``no glitchfind events'' refers to the residual with all events flagged by our software removed.}
    \label{fig:realppplot}
\end{figure*}
}

\newcommand{\injhistpplot}{
\begin{figure*}[t]
    \centering
    \includegraphics[width=0.9\textwidth]{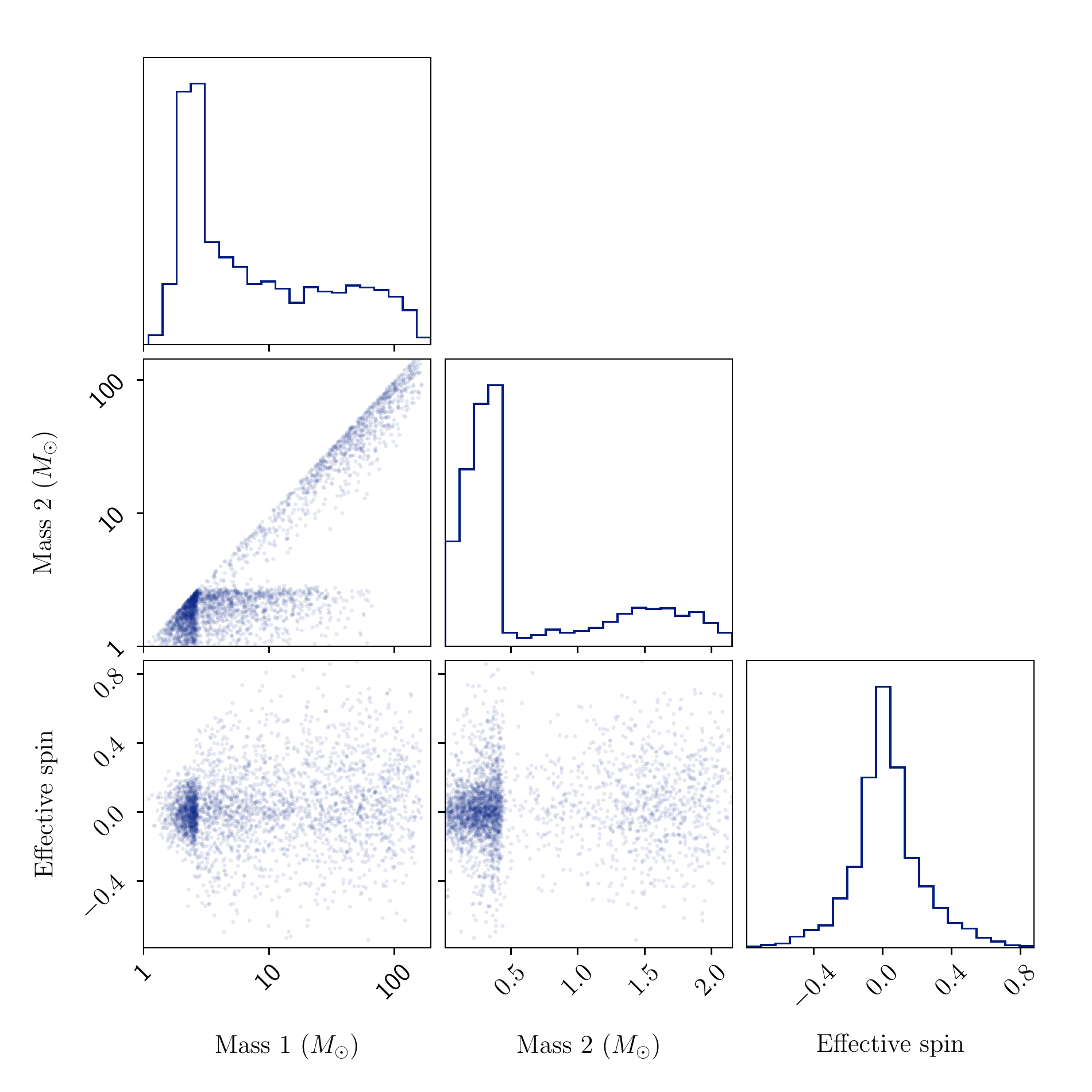}
    \caption{The distribution of masses and effective spins of the injected signals that were used in this study. The effective spin ($\chi_\text{eff}$) is a combination of the individual components' spin aligned with the orbital angular momentum of the system ($s_{1z}$ and $s_{2z}$) and the individual components' masses ($m_{1}$ and $m_{2}$). Specifically,  $\chi_\text{eff} = (m_1 s_{1z} + m_2 s_{2z})/(m_1 + m_2)$. Each panel corresponds to a histogram of the values for that parameter for all injections (the diagonal panels) or a visualization of two parameters (the off-diagonal panels), with each point corresponding to a single injection. All plotted injections were recovered with a false alarm rate less than $1 \mathrm{yr\,}^{-1}$.}
    \label{fig:injhistpplot}
\end{figure*}
}

\begin{document}

\title[Identifying glitches near gravitational-wave compact binary coalesces]{Identifying glitches near gravitational-wave signals from compact binary coalescences using the Q-transform}

\author{Leah Vazsonyi and Derek Davis}

\address{LIGO, California Institute of Technology, Pasadena, CA 91125, USA}
\ead{lvazsony@caltech.edu, dedavis@caltech.edu}
\date{today}

\begin{abstract}
We present a computational method to identify glitches in gravitational-wave data that occur nearby gravitational-wave signals from compact binary coalescences. The Q-transform, an established tool in LIGO-Virgo-KAGRA data analysis, computes the probability of any excess in the data surrounding a signal against the assumption of a Gaussian noise background, flagging any significant glitches. Subsequently, we perform validation tests on this computational method to ensure self-consistency in colored Gaussian noise, as well as data that contains a gravitational-wave event after subtracting the signal using the best-fit template. Finally, a comparison of our glitch identification results from real events in LIGO-Virgo's third observing run against the list of events which required glitch mitigation shows that this tool will be useful in providing precise information about data quality to improve astrophysical analyses of these events.


\end{abstract}

\section{Introduction}

Since the initial gravitational-wave detection GW150914, gravitational-wave interferometers have provided a new method of observing the universe through gravitational waves, having successfully observed a plethora of compact binary coalescence (CBC) signals so far~\cite{TheLIGOScientific:2014jea, VIRGO:2014yos, GW150914_paper, GWTC-1, GWTC-2, GWTC-2.1, GWTC-3}. These signals allow us to observe astrophysical phenomena through an entirely new spectrum beyond electromagnetic waves, providing the opportunity for ``multi-messenger astronomy,'' wherein we observe events using both gravitational and electromagnetic waves~\cite{GBM:2017lvd}.

Gravitational-wave interferometer data contains numerous noise artifacts, called ``glitches''~\cite{LIGO:2021ppb,Virgo:2022fxr,KAGRA:2020agh}.
These glitches can be caused by a variety of sources such as earthquakes~\cite{Schwartz:2020pso,Figura:2022btt}, lightning strikes~\cite{thunder_LIGO,Washimi:2021ogz}, or nearby human activity~\cite{AdvLIGO:2021oxw, Virgo:2022ypn}. 
Instrumental issues in the detectors themselves may also cause glitches~\cite{Accadia:2010zzb,Soni:2020rbu}.
If any of these glitches excite the detector at similar frequencies as binary coalescences, they may interfere with any true signal detected around that same time and bias analyses~\cite{Pankow:2018qpo, Powell:2018csz, Macas:2022afm, Payne:2022spz, Davis:glitch_sub}. Not only must we be able to sufficiently distinguish real gravitational-wave events from glitches, but also to understand the glitches well enough to recover portions of the gravitational-wave signal that may overlap with the glitch. 
Doing so will enable us to extract the gravitational-wave signal successfully, even when a loud glitch occurs nearby. 
For this reason, a variety of methods have been developed to subtract or otherwise mitigate glitches before analyzing gravitational-wave strain data~\cite{Cornish:2014kda,Wei:2019zlc,Zackay:2019kkv,Torres-Forne:2020eax,Cornish:2020dwh,Chatziioannou:2021ezd, Hourihane:2022doe,Merritt:2021xwh, Davis:glitch_sub}.

Following the third LIGO-Virgo observing run (O3)~\cite{GWTC-2, GWTC-2.1, GWTC-3}, improvements to the detector and data analysis tools allow us to be more sensitive to the power measured by the detector. This improved sensitivity means that the gravitational-wave signals will be louder and potentially clearer, but we also expect some noise artifacts to be louder as well~\cite{LIGO:2021ppb,Virgo:2022fxr,KAGRA:2020agh, 2018LRR....21....3A}. Thus, some previous glitch issues may be exacerbated in future observing runs.

In order to prevent glitches from biasing analyses of gravitational-wave signals, our first step is to identify if there is any excess power that needs to be mitigated. Generically, this excess power may occur due to a glitch or a signal, and it is sometimes not immediately clear which is causing the power. Although current methods exist to identify glitches~\cite{Robinet:2020lbf, Macleod:2021, kleinewelle}, none currently are able to quickly assign a significance of the presence of a glitch. This significance would assign a numerical value relating the the likelihood of such a signal, in our case under a null hypothesis of Gaussian noise. It is also common to identify glitches by visually inspecting images, which is a highly subjective manner of classification. Instead, it would be beneficial to statistically determine whether or not the noise is Gaussian; true glitches are caused by specific phenomena rather than the usual background in our observing runs, which follows a normal distribution. It also provides a more objective measure and definition for a glitch threshold. 

In this work, we introduce a new method to identify glitches using the Q-transform~\cite{Chatterji_thesis,Chatterji:2004qg} by rapidly computing a p-value, which measures the significance of any excess power in the data. With the assumption that the usual noise background is Gaussian, we can categorize any excess power which does not follow this expected distribution as a glitch and assign a p-value, assuming the noise to be Gaussian. In our work, we perform tests on this computation by testing its performance on (1) simulated gravitational-wave signals injected on a Gaussian background, (2) simulated gravitational-wave signals injected near a glitch, and (3) real gravitational-wave signals from O1, O2, and O3~\cite{GWTC-1, GWTC-2.1, GWTC-3} (LIGO observing runs 1, 2, and 3), some of which occur near a glitch. In each case, we also perform a significance test wherein we subtract the gravitational-wave signal from the data using a template generated by parameters recovered from that signal.

Our paper will proceed as follows:
Section \S \ref{sec:background} discusses methods to identify glitches, describing the mathematical background for computing p-values applicable to both gravitational-wave signals and glitches. This section includes a discussion on how we address clusters of tiles from the same glitch and how we account for the presence of an astrophysical signal in the data. 
Sections \S \ref{validation:gauss} and \S \ref{results:gauss} discuss the first set of validation tests in which we use a catalog of injections and recoveries to mimic gravitational-wave signal detections. We perform this test on a Gaussian background to check that we are able to recover real signals effectively under ideal conditions.
Sections \S \ref{validation:glitch} and \S \ref{results:glitch} discuss the second set of validation tests, which involves injecting a gravitational-wave signal near a loud glitch. In this case, we ensure that the tool accurately identifies the time periods containing both the signal and the glitch, and that the subtraction can recover the signal.
Section \S \ref{validation:real} and \S \ref{results:real} present our third validation test in which we use real signals from O3 and subtract the waveforms using the official recovery parameters for each event. This test is effectively a robustness check of our first test, but with a noisier background and no knowledge of the ``correct'' parameters of the true signal.
Finally, section \S \ref{conclusion} presents our concluding remarks, describes what we have learned from these tests, and discusses potential directions for future work.

\section{Glitch Identification Methods}\label{sec:background}
\subsection{Identifying Loud Tiles}
After measuring the gravitational-wave strain in the detector, we need some way of separating cases where excess power in the data corresponds to glitches or expected fluctuations from Gaussian noise. To do so, we analyze our data in the time-frequency domain, creating spectrograms generated using the Q-transform, a technique that has long been used in gravitational-wave data analysis to visualize data~\cite{Chatterji_thesis,Chatterji:2004qg,Blackburn:2008ah, GW150914_detchar, LIGO:2021ppb}.
The continuous Q-transform is given by~\cite{Chatterji_thesis,Chatterji:2004qg}
\begin{equation}
    x(\tau, f) = \int_{-\infty}^{+\infty} x(t) w(t-\tau) e^{-2\pi i f t} dt \text{ ,}
\end{equation}
where $w(t-\tau, f)$ is a window centered at a given time $\tau$. This window is dependent on the quality factor Q, defined as the frequency over the bandwidth, i.e.

\begin{equation}
    Q = \frac{f}{\delta f} \text{ .}
\end{equation}
However, gravitational-wave data is measured in discrete time samples and we hence use a discretized version of the Q-transform. This modified discretized transform is given by 
\begin{equation}
    x[m, k] = \sum_{n = 0}^{N-1} x[n] w[n-m, k] e^{-2\pi i n k / N} \text{ ,}
\end{equation}
creating a map of tiles, each of which corresponds to a specific time and frequency, where its value of interest is the energy at that tile. The set of tiles produced by this method is referred to as a ``Q-gram.'' In this work, we rely on the Q-transform and Q-gram implemented in \texttt{gwpy}~\cite{Macleod:2021}. As a preproccessing step, we whiten the data using the amplitude spectral density of the data, as estimated by \texttt{gwpy}. Analyzing whitened data simplifies the statistics introduced in the next section. We also will consider the tile energies, given by the square of the tile amplitudes, rather than simply the magnitudes, as this also reduces the complexity of the relevant distribution. We also normalize the tile energies by dividing by the median energy of all tiles in the spectrogram.

We note here that it is common to compute the discretized Q-transform tests for multiple values of Q within a given range, and the one with the largest individual tile energy is chosen to plot. We find later that this phenomenon affects the distribution of our p-values, so we subsequently fix Q. We select a value of 20 since this provides the best structure to our plots and allowed the windowing function to perform optimally for our purposes.

We therefore use this Q-transform that takes the strain data and computes the energy observed as a function of time and frequency. If there is a specific tile with more energy than the expected Gaussian background, we refer to this as a ``loud tile.'' These tiles could arise due to a gravitational-wave signal detection or a glitch. An example of a Q-gram and a spectrogram is shown in Fig.\ref{fig:spectrogram}.
This figure shows data from around both a gravitational wave signal and a glitch. 
Although the Q-gram and spectrogram visualizations are similar, the spectrogram interpolates the Q-gram data in order to estimate the energy at all time-frequency points. 
We use the measured energies of the individual tiles in a Q-gram throughout this work. 

\spectroplot

\subsection{Computing the P-value}

In order to determine whether the data contains a glitch or not, we assign a significance to each tile in our spectrogram. For stationary white noise, we expect these energies to follow an exponential distribution given by~\cite{Chatterji_thesis,Chatterji:2004qg} 
\begin{equation}\label{eq:probval}
    P(\epsilon > \epsilon[m, k]) \propto \mathrm{exp}(-\epsilon[m, k]/\mu_k) \text{ ,}
\end{equation}
where $k$ refers to the specific frequency and $\mu_k$ is the mean tile energy at the given frequency. The $m$ value is related to the time of the event from discretizing the continuous Q-transform.

We use this expected distribution to assign a probability, and hence significance, to each tile. This distribution is shown in Fig.~\ref{fig:distfig}. We fit the distribution of energies as suggested by Eq.~\ref{eq:probval}, but allow for some deviations from this theoretical prediction. We perform this fit according to Eq.~\ref{eq:probval}, ensuring that it is correctly normalized as well, according to 
\begin{equation}\label{eq:fit}
      P(\epsilon) = A e^{-\epsilon t} \text{ ,}
\end{equation}
where $\epsilon$ is the energy of a given tile and $A$ and $t$ are our fit parameters.

\distfig

To calculate the significance, we compute the p-value for the null hypothesis that the distribution of energies of the tiles in the Q-gram is consistent with the distribution expected from Gaussian noise. To start, we find the expected number of tiles above the energy of that tile given the size of the Q-gram and the fitted exponential model. We compute this by multiplying the size of the Q-gram by the integral of the probabilities that the energy is greater than or equal to the magnitude of that tile. This probability is given by 
\begin{equation} \label{probofe}
\begin{split}
    \tilde{P}(\epsilon > \epsilon_0) & = \int _{\epsilon_0} ^\infty P(\epsilon ') d \epsilon ' \\
    & = \int _{\epsilon_0} ^\infty \frac{A e^{-\epsilon  ' t}}{N} d \epsilon  ' \\
    & = \frac{A e^{-\epsilon_0 t}}{tN} \text{ .}
\end{split}
\end{equation}

Here, $\epsilon_0$ refers to some given threshold energy, $A$ and $t$ are our fit parameters, and $N$ is the number of tiles in the Q-gram. $\tilde{P}(\epsilon > \epsilon_0)$ is the probability that our Q-gram contains a tile above the threshold energy, and $P(\epsilon)$ is the normalized probability of a given energy given in Eqs.~\ref{eq:probval} and \ref{eq:fit}.

After computing the probability for each tile, we compute the probability of the Q-gram as a whole by assuming the same Gaussian distribution where the mean is the expected number of tiles. 

Assuming that the measured tile energies in Gaussian noise are a Poisson process, the probability of a observing at least one tile above a fixed energy is simply based on the rate of tiles above that energy and the total number of trials.
We use the probability computed according to Eq.~\ref{probofe} as the rate and the number of tiles in the Q-gram as the number of trials. Finally, we have that the probability of observing $\epsilon_0$ as the loudest tile in our Q-gram is based on the rate of occurrence, $\lambda$, and the length of data considered, $\tau$
\begin{equation}
\begin{split}
    P(\mathrm{Q\hbox{-}gram } \max{\epsilon} = \epsilon_0) &= 1 - e^{-\lambda \tau} \\
    &= 1 - \mathrm{exp}\left[ -\left( N\frac{A}{tN}e^{-\epsilon_0 t}\right) \left( N \right) \right] \\
    &= 1-\mathrm{exp}\left[ -\frac{AN}{t}e^{-\epsilon_0 t} \right] \text{ .}
\end{split}
\end{equation}

To find the glitch times, we first chose a p-value threshold and solve for the tile energy that has this probability of being observed. Any tiles with energies larger than this threshold energy are assumed to be due to non-Gaussian features in the data, i.e., glitches. Any times corresponding to those tiles are flagged as ``glitch times.''

\subsection{Identifying Clusters}

The computation has so far only considered each tile independently; in reality, if there are a number of significant tiles in a region, there may be other tiles that are part of the same glitch but are not statistically significant on their own. 
Thus, we include a clustering feature that includes these nearby loud tiles in the same data segment. 

To identify clusters, we first identify all tiles that meet some ``global'' p-value threshold. We then repeat the same analysis for a small time window around each of these tiles using a lower p-value threshold. 
For example, we can use 0.01 as the global p-value threshold but then use 0.2 as the second, lower threshold for clustering, which we use throughout our analysis. 
All tiles that meet this new threshold are included in the list of statically significant tiles.
The entire section of tiles is then categorized as a ``segment'' and flagged as a glitch.

\subsection{Subtracting the Signal}

 Since the test is only generically sensitive to excess power, this test would identify both glitches and astrophysical signals as loud tiles. 
 To ensure that any flagged tiles are non-astrophysical, it is possible to first subtract the astrophysical waveform and then analyze the residual. 
 To do this, we first need a reasonable estimate of the signal parameters from a different source.
 In practice, this test would be utilized after an event is first identified by a search algorithm, which analyzes the data with hundreds of thousands of templates to identify the best match.
 We can then use these best match parameters to subtract the signal. 

With the best fit parameters, care must be taken to ensure that the signal is precisely subtracted. 
In order to account for different signal processing methods or conventions between the analysis that identifies the signal parameters and our analysis, we use matched filtering to identify the time of the signal. 
We assume that the peak of the matched filter signal-to-noise (SNR) time series is the time of the signal we are trying to subtract. 
However, this introduces a different potential concern, as it is possible that the presence of a glitch in the data could bias the peak of the SNR time series. 
If we only assume that the time of the signal is during our analysis segment, it is possible that the peak of the SNR time series is actually a glitch that is not near the actual signal. 
We can address this by only considering peaks in a small time window around our initial estimate of the time of the signal. 
This still does not entirely remove this bias, as a glitch directly overlapping a signal could still bias the recovered time of arrival of the signal. 
However, in this case, the likely outcome is that we would still identify non-Gaussian features in the data, which could then be further investigated. 
To demonstrate how the importance of this timing information, we perform tests with and without enforcing that the peak of the SNR time series is within a small time window of our initial estimate.

\section{Validation Tests}
\subsection{Gaussian Tests}\label{validation:gauss}

Our initial set of tests represents an idealized form of recovering a gravitational-wave signal. We generate colored Gaussian noise with \texttt{Bilby}~\cite{Ashton:2018jfp} with the same power spectral density as a representative stretch of O3 data from LIGO Livingston in April 2019. This data was selected for a duration near an event with no data quality issues. Using this simulated time series, we inject a gravitational-wave signal, which we then seek to recover. Unless otherwise specified, we consider 8s of data around the relevant time in interest for all tests.

To approximate the process of recovering an astrophysical signal, we use the results of an injection campaign with \texttt{PyCBC}~\cite{pycbc_release, Usman:2015kfa}. 
For our simulated signal tests, we inject a waveform with the template parameters that were used to inject the signal in the \texttt{PyCBC} campaign but then attempt to subtract the signal using the parameters that \texttt{PyCBC} recovered from the injection.
Furthermore, we only consider injections that were identified by \texttt{PyCBC} with a false alarm rate of less than $1\text{yr}^{-1}$.
This process ensures that we are not assuming knowledge of the waveform parameters, which one does not have for astrophysical signals, and mirrors what would occur in practice for astrophysical signals. In subtracting the signal, we use a matched filter to find the peak SNR to determine the correct time of the subtraction.

Throughout this process, we generate three time series, each of which we analyze separately. First, there is a section of data that should simply approximate a Gaussian background, where we expect the p-values to be distributed as expected, i.e., 50\% have a p-value of 0.5 or lower, 90\% have a p-value of 0.9 or lower, etc. We also expect the tool to not identify time periods containing tiles with low p-values (we refer to these time segments as ``data quality flags''), which would indicate that this part of the data is inconsistent with Gaussian noise (due either to a glitch or a real event).
Next, we have the same portion of data but with a gravitational-wave signal injected, where we anticipate a very low p-value and a data quality flag across the duration of the waveform. The injected signal region should be considered significant.

If the first two time series are well-behaved (i.e., the background is Gaussian and the injection was successful without any artifacts), the final data portion will show how well we recovered the gravitational-wave signal. For a perfect recovery, we expect the time series to resemble the first portion containing only a Gaussian background. However, any mismatches in the manifestation of the gravitational-wave parameters of the injected and recovered signals will yield excess power in the data, so we expect some tiles to be louder and hence to potentially be flagged or have significant regions. Such regions should be within the duration of the injected waveform and should ideally be somewhat lower than in our injected time series. Thus, the distribution of p-values for many such tests should be distributed in a manner similar to that expected. However, one might anticipate a larger number of events at low p-value compared to the Gaussian data, indicating imperfect recovery of the signal by PyCBC.

We plot our results for the distribution of p-values in each test and compare it with the expected distribution (often referred to as a p-p plot).

\subsection{Glitches}\label{validation:glitch}

In our next set of tests, we work towards testing how well our computation performs at its primary objective: distinguishing a gravitational-wave signal and a glitch when both are present in the data. To test this, we take a set of glitch times classified by the Gravity Spy algorithm~\cite{Zevin:2016qwy, gravity_spy_dataset} from the second LIGO-Virgo observing run (O2) as our data and run a p-value computation and a data quality assessment. We select a set of the loudest ``blip'' glitches \cite{2019CQGra..36o5010C} from the site for simplicity, and since these glitches can be easily mistaken for gravitational wave signals. These glitches thus all have a signal-to-noise ratio above 10. Observing run 2 data were selected for the glitches since it was most readily available at the beginning of our work. We use a more recent PSD from the O3, but this choice does not bias our results. We then perform the same set of injections and subtractions described in section \S \ref{validation:gauss} to these glitches and repeat the p-value computation and data quality assessment at each stage. 
We randomize the time of the injection to occur within a small window of the glitch.
Injecting signals very near glitches allows us to test the ability of this tool to separate power from glitches and astrophysical signals, even when the two closely overlap. 

In this case, we expect the p-values to be extremely small, and the data quality flag should be present at the time of the glitch for all three time series. If the subtraction works correctly, we expect the data quality flag to be present for the injected signal but no longer be present once the subtraction has been performed. A comparison of the subtraction during these tests to Gaussian tests described in section \S \ref{validation:gauss} determines whether or not the glitch is interfering with our ability to recover the signal.

\subsection{Real Event Tests}\label{validation:real}

For our final tests, we analyze observed gravitational-wave events from O1, O2, and O3 and subtract a waveform corresponding to the reported source parameters. We consider all 90 events validated in GWTC-1~\cite{GWTC-1}, GWTC-2.1~\cite{GWTC-2.1}, and GWTC-3~\cite{GWTC-3}, which we collectively refer to as the GWTC, all of which had astrophysical probability greater than 0.5. The complete list of events we consider is listed in Tables~\ref{tab:glitch_tab_gwtc1}, \ref{tab:glitch_tab_gwtc2}, and \ref{tab:glitch_tab_gwtc3}, in \ref{app:results}. Such tests should mimic the injection and subtraction stages described in sections \S \ref{validation:gauss} and \S \ref{validation:glitch}, but would use real data and hence would not be as idealized. For example, the background is likely to be noisier and not as well-behaved as our re-sampled time series. Furthermore, the actual signals do not necessarily correspond perfectly to a template waveform, as was the case for the injected signals. We also expect that there will be glitches in the data, so this section may contain some subtracted time series with low p-values for this reason. We choose to use astrophysical probability as a threshold for inclusion in this section as this was the same threshold used in the GWTC analyses for investigating if glitches nearby events require mitigation.

\section{Results}

\subsection{Simulated Gaussian Tests}\label{results:gauss}

Since we expect the re-sampled data to be Gaussian and to behave as described in section \S \ref{sec:background}, the p-values for our simulated Gaussian data prior to injections or subtractions should be distributed as theoretically expected, i.e. a uniform distribution. Likewise, the distribution of p-values for injections should show an excess of low p-values and the subtracted results should be close to a uniform distribution. For the injected signals, we expect the p-values to be extremely low, except in the case of weak signals. Thus, we expect the distribution on our p-p plot (Fig. \ref{fig:realppplot}) to fall above that of the Gaussian tests, with a large excess of counts near zero. If all signals are loud, this should approach a horizontal line. For the subtraction time series, we expect the distribution to fall between the Gaussian background and the injected signals. The better the subtraction, the closer the subtraction distribution should be to the Gaussian background distribution.

The distributions of p-values that we measure for these tests are shown in Fig.~\ref{fig:gaussppplot}. 
We find that how Q is chosen for producing Q-grams does introduce a potential bias to our results; the left panel of Fig.~\ref{fig:gaussppplot} shows results with a fixed Q, while the right panel shows results when we test a range of Q values and choose the Q-gram with the highest individual tile energy. 
We quantify the level of agreement between our measured p-value distribution and the expected, uniform distribution using the Kolmogorov–Smirnov (KS) test~\cite{kolmogorov1933sulla}.

For tests with the Gaussian background data, we find that fixing Q results in a K-S test p-value of \red{0.21} compared to \red{0.06} when allowing Q to vary. 
For this reason, we focus on fixed-Q results in later sections. 
We also find that our results after subtracting the injected signal show larger deviations from the expected distribution than simply Gaussian noise, mostly at low p-values. 
Investigating cases where the p-value after subtraction was below 0.01, we find that these cases are primarily for high-SNR injections with parameters that are not well-captured by the aligned spin templates in the PyCBC template bank. 
The higher mismatch between the injected template and template used to subtract the signal resulted in residual excess power in the data after subtraction.

\gaussppplot

\subsection{Glitch Runs}\label{results:glitch}

As described in \S \ref{validation:glitch}, we expect the p-values for the initial data, the injection, and subsequent subtraction to be extremely low since a loud glitch will be present in all of these. Therefore, the p-values are not informative as to whether the tests behave correctly. 
Instead, we must check each of our runs to ensure that all three time series correctly flag the glitch, identifying it with a p-value less than 0.01. Furthermore, we expect a data quality flag on the injected signal, and for good recoveries (i.e., those that were successful in our simulated Gaussian tests), we expect the flag to no longer be present in the subtracted data. 

We find that the code always successfully returns a p-value less than 0.01, correctly flagging the glitch in all 300 cases. The injected data also always includes an appropriate data quality flag over the duration of the injected signal. 

Since the SNR of all glitches is higher than 10, the peak in the matched filter SNR time series is often the time of the glitch rather than the injected signal. To ensure that we are correctly subtracting the signal, we only look for peaks in the SNR time series within 0.1\,s of the estimated signal time (as reported by PyCBC). When this is done, we are able to correctly subtract the signal unless there is an almost perfect alignment of the signal and glitch. 
An example of data before and after subtraction is shown on the left side of Fig.~\ref{fig:glitch}.
If we instead considered the global peak in the SNR time series, we would often mistakenly attempt to subtract the glitch.
However, if the SNR of the injected signal were sufficiently loud, we would still be able to subtract the signal, as is shown on the left side of Fig.~\ref{fig:glitch}.
As expected, restricting the time window used to identify the SNR time series peak to be within a short time interval of the injected signal yields much better recovery rates than considering the global peak; we found that it was possible to subtract the signal perfectly and remove the data quality flag, even when the injection overlaps with the glitch.

\glitchfig

\subsection{Real Signals}\label{results:real}

As a final validation test, we analyze real gravitational-wave signals as described in section \S \ref{validation:real}. 
We analyze all detectors operating at the time of each event independently. 
Rather than use the same time window for all tests, as was done in previous sections, we use different time windows for each event that correspond to the same windows used in the GWTC to estimate the event source properties. 
These time windows ranged from 4 seconds to 128 seconds. 
In total, we analyze \red{209} different time series from \red{79} events.
The complete set of results, along with an additional discussion of the results, can be found in \ref{app:results}.

We repeat the same test of the p-p plot as before (Sec. \ref{results:gauss}), shown in Fig.~\ref{fig:realppplot}. 
This figure includes the results using the data before subtraction and the residual after subtraction. 
As data around these events is known to contain glitches, we also attempt to remove events that contain glitches. 
To do this, we use a p-value threshold of 0.01 to flag glitches; we then plot the distribution of results with these events excluded. 
\red{33} of the 209 total time series were flagged with this threshold. 
Visual inspection of the spectrograms from flagged time series confirmed that glitch was visible in all cases. 
To account for any events that would be falsely flagged as glitches, when plotting the measured p-value distribution, we assume that 1\% of the results are still below 0.01. 
After removing these events, we find that the distribution of p-values is consistent with Gaussian noise at a \red{K-S test p-value of 0.037}. 
Although this may indicate some remaining excess power that is not Gaussian, it is not possible to identify any additional specific events that are not consistent with Gaussian noise. 

As an additional comparison, we instead only remove events that required glitch subtraction in the GWTC from our results.
This requirement removes \red{18} time series from our results. 
The distribution of p-values after removing these events is also shown in Fig.~\ref{fig:realppplot}.
We find that even after removing these events, we still have an excess of events with p-value less than 0.01, suggesting that there are still significant non-Gaussian features in the remaining time series that were not subtracted.
As the metrics used to decide if glitch mitigation was required in these analyses are different than simply identifying if a glitch is present~\cite{Davis:glitch_sub}, it is possible that some glitches in the data were not deemed to require subtraction. 
This may explain why these cases were not mitigated in the GWTC analyses. 

Additional visual inspection of time series was completed for cases where this algorithm disagreed with the list of glitches subtracted from the GWTC. 
Any time series with a p-value of less than 0.01 that did not use glitch mitigation or a time series with a p-value of more than 0.01 that did use glitch mitigation was investigated. 
In all cases where this algorithm identifies a glitch, excess power was visible in the produced spectrogram. 
Many of these flagged glitches were short in duration at high frequency and did not overlap the signal.
Conversely, no excess power was visible in cases where the algorithm did not identify a glitch. 
In the majority of these cases, the glitch that was subtracted was out of the time-frequency region considered by this tool.
Although these checks did identify multiple cases where excess power is visible in the data that was not subtracted before analyses, it is not known if these glitches could bias parameter estimation results. 
Additional investigations, similar to~\cite{Davis:glitch_sub}, would be required to understand any potential biases and are out of the scope of this work. 
However, the additional glitches we found with this tool demonstrate that this type of algorithmic approach to glitch identification is more accurate at identifying glitches than only visual inspection of the data. 
Additional comparisons between our results and those of the GWTC can be found in \ref{app:results}.

\realppplot

\section{Conclusion}\label{conclusion}

In this work, we have presented and tested a method for identifying glitches that may occur near gravitational-wave signals. The method rapidly computes whether a statistically significant glitch occurred within a section of data, flags noisy regions, and then subtracts the CBC  gravitational-wave signal to effectively separate the signal and the glitch. Although the exact time to run the code varies depending on the length and complexity of the time series, it generally takes less than a few minutes for the examples shown in this work.

We have shown that this tool's significance computation and data quality flagging were functioning correctly for a Gaussian background, gravitational-wave signal, and a loud glitch. We found that the measured p-values for all cases are distributed as expected and minimally biased: Gaussian data follows the expected distribution, and all loud events (including both gravitational-wave signals and glitches) yield a small p-value. Matching a template waveform and subsequently subtracting it from the gravitational-wave data leaves a p-value somewhat higher than the Gaussian background case due to template mismatches with the actual signal present in the data.
Our tests injecting a signal near a glitch and subsequently subtracting a template waveform showed significantly more success when there is separate knowledge of the time of arrival of the gravitational-wave signal. Still, it is often possible to clearly identify glitches even without such additional information. 

One of the main potential use cases is to automate glitch detection in cases where this is currently done by visual inspection of spectrograms. 
We compared the glitch identification results based on this algorithm against the list of O1, O2, and O3 events that required glitch mitigation~\cite{GWTC-1, GWTC-2.1, GWTC-3}. 
In general, we find agreement between these two methods of glitch identification, but this algorithm identifies multiple additional glitches in the data surrounding events that were not mitigated. 
It is important to note that all glitches near events may not require mitigation; this reason may explain some of the differences we identify. 
As this algorithm is not able to evaluate the potential impact of a glitch on an astrophysical analysis, additional methods would be needed to answer this question.

This algorithm requires only seconds to run, making it potentially useful to aid in investigations of gravitational-wave events that are detected in low-latency.
Glitches that overlap gravitational-wave signals may impact the estimate of parameters that are important to multi-messenger astronomy, such as the distance and sky position of the source~\cite{Macas:2022afm}. 
The faster that such a bias is detected will reduce the chance that resources are wasted following up incorrect information and increase the speed at which unbiased results can be released. 

An important limitation of this method is that signal subtraction is only possible when the parameters of the event are well known. In the case of CBC signals, we have shown that it is generally sufficient to use the template parameters from the search that identified the signal. For signals that are detected by unmodelled searches~\cite{Klimenko:2015ypf, Lynch:2015yin}, separating astrophysical excess power from instrumental excess power is more difficult. If information such as the time of arrival, duration, and bandwidth of a signal is known, it is possible to exclude that time-frequency region before calculating this statistic. However, it is not possible for this method to differentiate an unmodelled astrophysical signal that directly overlaps a glitch. 
For a CBC signal, our tool could be tuned to look only in the time-frequency regime near the signal.

We hope that this tool can be used in future observing runs to increase the speed, accuracy, and automation of glitch identification. 
As the rate of gravitational-wave detection increases, we expect that this method, and similar statistical tools, will be an important component of adapting gravitational-wave analysis methods to handle large numbers of events. 
By relying on these types of statistical methods, personpower that was previously spent on visual inspection of the data can be diverted to other applications. 
Furthermore, the time-frequency information about identified glitches that this tool generates can be used to further automated additional gravitational-wave analyses, such as glitch subtraction, which are important components of arriving at accurate astrophysical conclusions.  

\section{Acknowledgements}
The authors would like to thank the LIGO-Virgo-KAGRA detector characterization groups for their input and suggestions during the development of this work. 
We thank Siddharth Soni for his comments during internal review of this manuscript.
LV and DD are supported by the NSF as a part of the LIGO Laboratory.

This research has made use of data or software obtained from the Gravitational Wave Open Science Center (gw-openscience.org), a service of LIGO Laboratory, the LIGO Scientific Collaboration, the Virgo Collaboration, and KAGRA. LIGO Laboratory and Advanced LIGO are funded by the United States National Science Foundation (NSF) as well as the Science and Technology Facilities Council (STFC) of the United Kingdom, the Max-Planck-Society (MPS), and the State of Niedersachsen/Germany for support of the construction of Advanced LIGO and construction and operation of the GEO600 detector. Additional support for Advanced LIGO was provided by the Australian Research Council. Virgo is funded, through the European Gravitational Observatory (EGO), by the French Centre National de Recherche Scientifique (CNRS), the Italian Istituto Nazionale di Fisica Nucleare (INFN) and the Dutch Nikhef, with contributions by institutions from Belgium, Germany, Greece, Hungary, Ireland, Japan, Monaco, Poland, Portugal, Spain. The construction and operation of KAGRA are funded by Ministry of Education, Culture, Sports, Science and Technology (MEXT), and Japan Society for the Promotion of Science (JSPS), National Research Foundation (NRF) and Ministry of Science and ICT (MSIT) in Korea, Academia Sinica (AS) and the Ministry of Science and Technology (MoST) in Taiwan.

This material is based upon work supported by NSF’s LIGO Laboratory 
which is a major facility fully funded by the 
National Science Foundation.
LIGO was constructed by the California Institute of Technology 
and Massachusetts Institute of Technology with funding from 
the National Science Foundation, 
and operates under cooperative agreement PHY-1764464. 
Advanced LIGO was built under award PHY-0823459.
The authors are grateful for computational resources provided by the 
LIGO Laboratory and supported by 
National Science Foundation Grants PHY-0757058 and PHY-0823459.
This work carries LIGO document number P2200223.

\appendix

\section{Complete GWTC results}\label{app:results}

In this appendix, we present complete details of our results for GWTC events~\cite{GWTC-1, GWTC-2.1, GWTC-3}.
These measured p-values and analysis configurations for each event can be found in Tables~\ref{tab:glitch_tab_gwtc1}, ~\ref{tab:glitch_tab_gwtc2}, and \ref{tab:glitch_tab_gwtc3}.
To compare our results with those in the GWTC, we first identify glitches using a p-value threshold. 
If the measured p-value is less than 0.01, we consider this a glitch. 
We find that \red{39} of the analyzed time series meet this threshold. 
Conversely, only \red{19} time series used glitch mitigation in the GWTC analyses. 

In total, we identify 28 differences between our analysis and the GWTC. 
We find 24 glitches that are not listed in the GWTC, while we do not find 4 of the glitches listed in the GWTC.
It is possible that these glitches were identified as a part of the GWTC analyses but not flagged for glitch mitigation, as additional metrics used to decide if glitch mitigation was required~\cite{Davis:glitch_sub}.
As mentioned in the main text, many of the glitches identified by this algorithm are short in duration and high frequency, which likely did not require mitigation. 
However, a significant fraction of the glitches we identify are clearly visible in a spectrogram of the data. 
Of the 4 cases where we do not identify a glitch, 3 of these where due to the frequency limits used in our analysis. If we lowered the lower frequency limit to 10\,Hz, these glitches would have been identified by our analysis as significant. 
The level of agreement between our results and those from the GWTC do appear to be detector-dependent.
We find 99\% agreement (83/84) with the results for LIGO Hanford, 87\% agreement (76/87) with the results for LIGO Livingston, and 75\% agreement (49/65) with the results for Virgo.  
If we only consider cases where this algorithm identifies a glitch, 
the agreement rate is 80\% (4/5) for LIGO Hanford, 59\% (10/17) for LIGO Livingston, and 6\% (1/17) for Virgo. 

The reasons for these detector-dependent rates may be due to a combination of factors, including the rate of glitches in each detector, the types of glitches present in each detector, the SNR of signals in each detector, and different interpretations of the data from person to person during the visual inspection process.
The majority of glitches that this tool identified that were not mitigated in GWTC analyses were short in duration and relatively low normalized energy ($\approx$20-25). 
These cases are likely to have simply been missed by only visual inspection. 
However, in a number of cases, our tool identifies high normalized energy ($> 30$) tiles from glitches that were not mitigated. 
Examples include GW200216\_220804 and GW200224\_222234 at LIGO Livingston as well as GW200112\_155838 and GW200129\_065458 at Virgo. 
These glitches appear similar to scattered light glitches that were common at all sites during O3. 
Such cases demonstrate the benefits of this algorithmic approach to glitch identification; the large number of events surveyed during these analyses makes it likely that visual inspection methods will not be applied consistently to all events, leading to cases where glitches are not identified. 

\begin{table}[p]
\footnotesize
\begin{tabularx}{\linewidth}{l l r r r X l}
Event name & Window & H1 p-value & L1 p-value &  V1 p-value & GWTC result & Agree?\\
\hline
GW150914 & 4\,s & 0.490 & 0.624 & - & - & \cmark \\
GW151012 & 4\,s & 0.167 & 0.192 & - & - & \cmark \\
GW151226 & 4\,s & 0.999 & 0.789 & - & - & \cmark \\
GW170104 & 4\,s & 0.606 & 0.107 & - & - & \cmark \\
GW170608 & 4\,s & 0.426 & \textbf{0.007} & - & - & \xmark \\
GW170729 & 4\,s & 0.282 & 0.012 & 0.091 & - & \cmark \\
GW170809 & 4\,s & 0.591 & 0.121 & \textbf{0.006} & - & \xmark \\
GW170814 & 4\,s & 0.300 & 0.867 & \textbf{0.002} & - & \xmark \\
GW170817 & 64\,s & 0.472 & \textbf{$<$ 0.001} & \textbf{$<$ 0.001} & L1 glitch & \xcmark \\
GW170818 & 4\,s & 0.688 & 0.101 & \textbf{0.001} & - & \xmark \\
GW170823 & 4\,s & 0.950 & 0.086 & - & - & \cmark \\
\end{tabularx}
\caption{\footnotesize The results of our residual test for all events in GWTC-1. The measured p-value is listed separately for each detector; cases where a detector was not operating are denoted as $-$. All results with a p-value less than 0.01 are considered impacted by glitches. For each of event, a \cmark means we find the same glitches as GWTC-1, a \xmark means our list of glitches completely disagrees with GWTC-1 and \xcmark indicates partial agreement. We generally find agreement between the glitches found by this method and those listed in GWTC-1, but multiple differences are noted.}
\label{tab:glitch_tab_gwtc1}
\end{table}

\begin{table}[p]
\footnotesize
\begin{tabularx}{\linewidth}{l l r r r X l}
Event name & Window & H1 p-value & L1 p-value &  V1 p-value & GWTC result & Agree?\\
\hline
GW190403\_051519 & 4\,s & 0.950 & 0.950 & 0.946 & - & \cmark \\
GW190408\_181802 & 8\,s & 0.516 & 0.131 & 0.804 & - & \cmark \\
GW190412 & 8\,s & 0.355 & 0.036 & 0.589 & - & \cmark \\
GW190413\_052954 & 4\,s & 0.011 & 0.895 & 0.665 & - & \cmark \\
GW190413\_134308 & 4\,s & 0.540 & \textbf{$<$ 0.001} & 0.970 & L1 glitch & \cmark \\
GW190421\_213856 & 4\,s & 0.172 & 0.176 & - & - & \cmark \\
GW190425 & 128\,s & - & \textbf{$<$ 0.001} & \textbf{$<$ 0.001} & L1 glitch & \xcmark \\
GW190426\_190642 & 4\,s & 0.707 & 0.169 & 0.744 & - & \cmark \\
GW190503\_185404 & 4\,s & 0.359 & \textbf{0.002} & \textbf{0.001} & L1 glitch & \xcmark \\
GW190512\_180714 & 8\,s & 0.660 & 0.455 & 0.198 & - & \cmark \\
GW190513\_205428 & 4\,s & 0.908 & 0.165 & 0.335 & L1 glitch & \xmark \\
GW190514\_065416 & 4\,s & 0.011 & 0.948 & - & L1 glitch & \xmark \\
GW190517\_055101 & 4\,s & 0.599 & 0.155 & 0.137 & - & \cmark \\
GW190519\_153544 & 4\,s & 0.475 & 0.862 & 0.562 & - & \cmark \\
GW190521 & 4\,s & 0.343 & 0.130 & 0.353 & - & \cmark \\
GW190521\_074359 & 4\,s & 0.207 & \textbf{0.009} & - & - & \xmark \\
GW190527\_092055 & 8\,s & 0.997 & 0.958 & - & - & \cmark \\
GW190602\_175927 & 4\,s & 0.174 & \textbf{$<$ 0.001} & 0.684 & - & \xmark \\
GW190620\_030421 & 4\,s & - & 0.499 & 0.989 & - & \cmark \\
GW190630\_185205 & 4\,s & - & 0.418 & 0.391 & - & \cmark \\
GW190701\_203306 & 4\,s & 0.772 & \textbf{0.004} & 0.348 & L1 glitch & \cmark \\
GW190706\_222641 & 4\,s & 0.025 & 0.203 & 0.269 & - & \cmark \\
GW190707\_093326 & 16\,s & 0.045 & 0.439 & - & - & \cmark \\
GW190708\_232457 & 8\,s & - & 0.647 & 0.171 & - & \cmark \\
GW190719\_215514 & 4\,s & 1.000 & 0.720 & - & - & \cmark \\
GW190720\_000836 & 16\,s & 0.021 & 0.020 & 0.045 & - & \cmark \\
GW190725\_174728 & 16\,s & 0.085 & 0.739 & 0.150 & - & \cmark \\
GW190727\_060333 & 4\,s & 0.701 & 0.864 & 0.311 & L1 glitch & \xmark \\
GW190728\_064510 & 16\,s & 0.418 & 0.463 & 0.488 & - & \cmark \\
GW190731\_140936 & 4\,s & 0.722 & 0.245 & - & - & \cmark \\
GW190803\_022701 & 4\,s & 0.046 & 0.624 & 0.844 & - & \cmark \\
GW190805\_211137 & 4\,s & 0.468 & 0.302 & 0.970 & - & \cmark \\
GW190814 & 32\,s & \textbf{0.003} & \textbf{$<$ 0.001} & 0.067 & L1 glitch & \xcmark \\
GW190828\_063405 & 4\,s & 0.227 & 0.755 & \textbf{0.004} & - & \xmark \\
GW190828\_065509 & 8\,s & 0.089 & 0.672 & \textbf{0.008} & - & \xmark \\
GW190910\_112807 & 4\,s & - & 0.918 & 0.661 & - & \cmark \\
GW190915\_235702 & 8\,s & 0.320 & 0.822 & 0.984 & - & \cmark \\
GW190916\_200658 & 8\,s & 0.727 & 0.645 & 0.877 & - & \cmark \\
GW190917\_114630 & 64\,s & 0.146 & 0.103 & 0.117 & - & \cmark \\
GW190924\_021846 & 32\,s & 0.989 & \textbf{$<$ 0.001} & \textbf{0.009} & L1 glitch & \xcmark \\
GW190925\_232845 & 8\,s & 0.342 & - & \textbf{$<$ 0.001} & - & \xmark \\
GW190926\_050336 & 4\,s & 0.986 & 0.684 & 0.142 & - & \cmark \\
GW190929\_012149 & 4\,s & 0.637 & 0.059 & 0.224 & - & \cmark \\
GW190930\_133541 & 16\,s & 0.316 & 0.025 & - & - & \cmark \\
\end{tabularx}
\caption{\footnotesize The results of our residual test for all events in GWTC-2.1. The measured p-value is listed separately for each detector; cases where a detector was not operating are denoted as $-$. All results with a p-value less than 0.01 are considered impacted by glitches. For each of event, a \cmark means we find the same glitches as GWTC-2.1, a \xmark means our list of glitches completely disagrees with GWTC-2.1 and \xcmark indicates partial agreement. We generally find agreement between the glitches found by this method and those listed in GWTC-2.1, but multiple differences are noted.}
\label{tab:glitch_tab_gwtc2}
\end{table}

\begin{table}[p]
\footnotesize
\begin{tabularx}{\linewidth}{l l r r r X l}
Event name & Window & H1 p-value & L1 p-value &  V1 p-value & GWTC result & Agree?\\
\hline
GW191103\_012549 & 16\,s & 0.615 & 0.772 & - & - & \cmark \\
GW191105\_143521 & 16\,s & 0.098 & \textbf{0.004} & \textbf{$<$ 0.001} & V1 glitch & \xcmark \\
GW191109\_010717 & 4\,s & \textbf{$<$ 0.001} & 0.158 & - & H1 and L1 glitch & \xcmark \\
GW191113\_071753 & 64\,s & \textbf{$<$ 0.001} & 0.166 & \textbf{$<$ 0.001} & H1 glitch & \xcmark \\
GW191126\_115259 & 64\,s & 0.661 & \textbf{$<$ 0.001} & - & - & \xmark \\
GW191127\_050227 & 8\,s & \textbf{$<$ 0.001} & 0.500 & 0.765 & H1 glitch & \cmark \\
GW191129\_134029 & 16\,s & 0.010 & 0.878 & - & - & \xmark \\
GW191204\_110529 & 8\,s & 0.670 & 0.025 & - & - & \cmark \\
GW191204\_171526 & 8\,s & 0.986 & 0.483 & - & - & \cmark \\
GW191215\_223052 & 8\,s & 0.021 & 0.492 & \textbf{$<$ 0.001} & - & \xmark \\
GW191216\_213338 & 16\,s & 0.601 & - & 0.025 & - & \cmark \\
GW191219\_163120 & 32\,s & \textbf{$<$ 0.001} & \textbf{$<$ 0.001} & \textbf{0.006} & H1 and L1 glitch & \xcmark \\
GW191222\_033537 & 8\,s & 0.721 & 0.136 & - & - & \cmark \\
GW191230\_180458 & 4\,s & 0.299 & 0.241 & 0.350 & - & \cmark \\
GW200112\_155838 & 4\,s & - & 0.475 & \textbf{$<$ 0.001} & - & \xmark \\
GW200115\_042309 & 64\,s & 0.466 & \textbf{$<$ 0.001} & 0.567 & L1 glitch & \cmark \\
GW200128\_022011 & 4\,s & 0.954 & 0.721 & - & - & \cmark \\
GW200129\_065458 & 8\,s & 0.159 & \textbf{$<$ 0.001} & \textbf{$<$ 0.001} & L1 glitch & \xcmark \\
GW200202\_154313 & 16\,s & 0.358 & 0.500 & 0.010 & - & \cmark \\
GW200208\_130117 & 4\,s & 0.963 & 0.951 & 0.022 & - & \cmark \\
GW200208\_222617 & 8\,s & 0.362 & 0.832 & 0.926 & - & \cmark \\
GW200209\_085452 & 4\,s & 0.567 & 0.315 & 0.962 & - & \cmark \\
GW200210\_092254 & 16\,s & 0.319 & 0.356 & 0.098 & - & \cmark \\
GW200216\_220804 & 4\,s & 0.265 & \textbf{$<$ 0.001} & 0.822 & - & \xmark \\
GW200219\_094415 & 4\,s & 0.118 & 0.668 & 0.023 & - & \cmark \\
GW200220\_061928 & 4\,s & 0.195 & 0.279 & 0.410 & - & \cmark \\
GW200220\_124850 & 4\,s & 0.894 & 0.627 & - & - & \cmark \\
GW200224\_222234 & 4\,s & 0.267 & \textbf{$<$ 0.001} & 0.881 & - & \xmark \\
GW200225\_060421 & 8\,s & 0.413 & 0.169 & - & - & \cmark \\
GW200302\_015811 & 8\,s & 0.027 & - & 0.969 & - & \cmark \\
GW200306\_093714 & 16\,s & 0.014 & 0.395 & - & - & \cmark \\
GW200308\_173609 & 16\,s & 0.693 & 0.983 & 0.158 & - & \cmark \\
GW200311\_115853 & 4\,s & 0.027 & 0.547 & 0.199 & - & \cmark \\
GW200316\_215756 & 16\,s & 0.194 & 0.194 & \textbf{0.006} & - & \xmark \\
GW200322\_091133 & 16\,s & 0.292 & 0.898 & 0.166 & - & \cmark \\
\end{tabularx}
\caption{\footnotesize The results of our residual test for all events in GWTC-3. The measured p-value is listed separately for each detector; cases where a detector was not operating are denoted as $-$. All results with a p-value less than 0.01 are considered impacted by glitches. For each of event, a \cmark means we find the same glitches as GWTC-3, a \xmark means our list of glitches completely disagrees with GWTC-3 and \xcmark indicates partial agreement. We generally find agreement between the glitches found by this method and those listed in GWTC-3, but multiple differences are noted.}
\label{tab:glitch_tab_gwtc3}
\end{table}

\section{Details of injection campaign}\label{app:inject}

In order to access the effectiveness of this tool with realistic mismatches between the estimated signal parameters and the true parameters, we carried out an injection campaign with the PyCBC search algorithm~\cite{Usman:2015kfa}.
We estimated the significance of 6000 injections and identified 2963 of these with a false alarm rate of less than $1\text{yr\,}^{-1}$. Only injections with a false alarm rate less than this threshold were considered in the analysis. This cut was applied to ensure that only injections that were unambiguously identified were considered in the analysis. Figure~\ref{fig:injhistpplot} shows the distribution of the mass and spin parameters of the injections with false alarm rate of less than $1\text{yr\,}^{-1}$. The distribution of all injections and this subset of injections are qualitatively the same, with no significant biases in what type of signals were recovered. 

\injhistpplot

\section*{References}

\bibliographystyle{unsrt}
\bibliography{references}

\end{document}